\documentclass[a4paper]{PoS}
\usepackage{amsbsy}		
\usepackage{amsfonts}		
\usepackage{amsmath}		
\usepackage{cite}
\usepackage{mathrsfs}
\usepackage{dsfont}

\let\Re\relax
\let\Im\relax

\DeclareMathOperator{\Re}{Re}
\DeclareMathOperator{\Im}{Im}

\newenvironment{Eqnarray}%
         {\arraycolsep 0.14em\begin{eqnarray}}{\end{eqnarray}}
\def\beqa{\begin{Eqnarray}}
\def\eeqa{\end{Eqnarray}}
\def\beq{\begin{equation}}
\def\eeq{\end{equation}}
\def\eq#1{eq.~(\ref{#1})}

\def\eqs#1#2{eqs.~(\ref{#1}) and (\ref{#2})}

\def\eqs#1#2{eqs.~(\ref{#1}) and (\ref{#2})}

\def\vev#1{\langle #1 \rangle}
\def\half{\tfrac12}
\def\quarter{\tfrac14}
\def\phm{\phantom{-}}
\def\nn{\nonumber}
\def\cbma{c_{\beta-\alpha}}
\def\sbma{s_{\beta-\alpha}}
\def\cba{\cos(\beta-\alpha)}
\def\sba{\sin(\beta-\alpha)}

\def\mhh{m_{H}}
\def\mhl{m_{h}}

\def\anti{\overline}

\def\ur{U_R}
\def\dr{D_R}

\def\lsim{\mathrel{\raise.3ex\hbox{$<$\kern-.75em\lower1ex\hbox{$\sim$}}}}
\def\gsim{\mathrel{\raise.3ex\hbox{$>$\kern-.75em\lower1ex\hbox{$\sim$}}}}
\makeatletter

\newcommand{\Rmnum}[1]{\expandafter\@slowromancap\romannumeral #1@}
\makeatother

\title{Future Higgs Studies: A Theorist's Outlook}
\ShortTitle{Future Higgs Studies: A Theorist's Outlook}

\author{\speaker{Howard E.~Haber}%
        \\
       Santa Cruz Institute for Particle Physics\\
       University of California, Santa Cruz, CA 95064 USA \\
       E-mail: \email{haber@scipp.ucsc.edu}}

\abstract{We examine some of the theoretical and phenomenological implications of the Higgs boson discovery and discuss
what these imply for future Higgs studies at the LHC and future colliders.  In particular, one of the 
outstanding unanswered questions is whether additional scalars beyond the observed Higgs boson 
are present in the spectrum of fundamental particles.   Any theory of a non-minimal Higgs sector must
possess a scalar state whose properties are approximately those of the Standard Model Higgs boson.
This can be achieved in the so-called alignment limit of the extended scalar sector.  Examples of scalar sectors in which
an approximate alignment limit is achieved are surveyed.
}

\FullConference{Prospects for Charged Higgs Discovery at Colliders\\
                 3-6 October 2016\\
                 Uppsala, Sweden}

\begin{document}

\section{Introduction}

The discovery of  the Higgs boson in 2012~\cite{Aad:2012tfa,Chatrchyan:2012xdj}
 marked an important milestone in the study of fundamental particles and their interactions.
The Standard Model (SM) of particle physics is now complete.  Indeed, there are no definitive departures from the Standard Model observed in experiments conducted at high energy collider facilities.  
Nevertheless, some fundamental microscopic phenomena must necessarily lie outside of the purview of the SM.
These include: neutrinos with non-zero mass~\cite{numass}; dark matter~\cite{darkmatter};  the baryon asymmetry of the universe~\cite{White:2016nbo};  the suppression of CP-violation in the strong interactions (the so-called strong CP problem~\cite{Kim:2008hd});  gauge coupling unification~\cite{guts}; inflation in the early universe~\cite{inflation}; dark energy~\cite{darkenergy}; and the gravitational interaction.  None of these phenomena can be explained within the framework of the SM alone.

As a result, the SM should be regarded at best as a low-energy effective field theory, which is valid below some high energy scale $\Lambda$.   For example, a credible theory of neutrino masses (e.g., the type-I seesaw model~\cite{numass}) posits the existence of a right-handed electroweak singlet Majorana neutrino of mass of order $10^{14}~{\rm GeV}$.   The gravitational force is governed by Planck-scale physics corresponding to $\Lambda\sim 10^{19}$~GeV.  
Henceforth, we shall define $\Lambda$ to be the lowest energy scale at which the SM breaks down.  The predictions made by the SM depend on a number of parameters that must be taken as input to the theory.   These parameters are 
sensitive to ultraviolet (UV) physics, and since the physics at very high energies is not known, one cannot predict their values.
In general, fermions and boson masses depend differently on $\Lambda$~\cite{Weisskopf:1939zz}.
On the one hand, fermion masses are
logarithmically sensitive to UV physics, due to
the chiral symmetry of massless fermions, i.e.
$
\delta m_F\sim m_f\ln(\Lambda^2/m_F^2)$.
In contrast, no such symmetry exists to protect masses of spin-0 bosons (in the absence of
supersymmetry), and consequently we expect quadratic
sensitivity of the scalar boson squared mass to UV physics,
$
\delta m_B^2\sim \Lambda^2$.

In the SM, the Higgs scalar potential, 
\beq \label{V}
V(\Phi)=-\mu^2(\Phi^\dagger\Phi)+\tfrac12\lambda(\Phi^\dagger\Phi)^2\,,
\eeq
where $\mu^2=\frac12\lambda v^2$ depends on the vacuum expectation value (vev) $v$ of the Higgs field.  
The parameter $\mu^2$ is quadratically sensitive to $\Lambda$.  Hence, to obtain $v\simeq 246$~GeV in a theory
where $v\ll \Lambda$ requires a significant fine-tuning of the ultraviolet parameters of the fundamental theory.
Indeed, the one-loop contribution to the squared mass parameter $\mu^2$ would be expected to be of
order $(g^2/16\pi^2)\Lambda^2$.  Setting this quantity to be of order of $v^2$ (to avoid an \textit{unnatural} cancellation
between the tree-level parameter and the loop corrections) yields
$
\Lambda\simeq 4\pi v/g\sim {O}(1~{\rm TeV})
$.
A \textit{natural} theory of electroweak symmetry breaking (EWSB) would seem to require new physics at the TeV scale associated with the EWSB dynamics.  

There have been a number of theoretical proposals to explain the origin of the EWSB energy scale:
(1) naturalness is restored by supersymmetry which ties the bosons to
the more well-behaved fermions~\cite{susy}; (2) the Higgs boson is an approximate Goldstone boson, the only other
known mechanism for keeping an elementary scalar light~\cite{dewsb}; (3) The Higgs boson is a composite scalar, with an inverse length of
order the TeV-scale~\cite{dewsb}; (4) the EWSB scale is chosen by some vacuum selection mechanism{~\cite{Agrawal:1998xa}.
Of course, maybe none of these explanations are relevant, and the EWSB energy scale (which appears to us to be highly fine-tuned) is simply the result of some initial condition whose origin will never be discernible.

In light of these remarks, how do we make further progress?   We have at our disposal a very successful experimental particle physics facility---the Large Hadron Collider (LHC), which has only begun a comprehensive probe of the TeV-energy scale.  To the experimentalists, I say: ``keep searching for new physics beyond the SM (BSM).''   Any observed departures from SM predictions will contain critical clues to a more fundamental theory of elementary particles and their interactions.  To the theorists, I say: ``find new examples of  BSM physics (which might provide a natural explanation to the EWSB scale) that may have been overlooked in LHC searches.''  But what if no signals for BSM physics emerge soon?  My answer is: ``look to the Higgs sector.''  After all, we have only recently discovered a most
remarkable particle that seems to be like nothing that
has ever been seen before---an elementary scalar boson.
Shouldn't we probe this state thoroughly 
and explore its properties with as much precision as possible?

Putting considerations of naturalness aside, two critical questions to be addressed in future LHC experimentation are:

1. Are there additional Higgs bosons to be discovered? (This includes new charged scalars of interest to this conference.) If fermionic
matter and the gauge sector of the SM are non-minimal, why shouldn't scalar matter also be
non-minimal? To
paraphrase I.I. Rabi, ``who ordered that?''}

2. If we measure the Higgs properties with sufficient precision,
will deviations from SM-like Higgs behavior be revealed?


One might be concerned that adding additional Higgs scalars to the theory will exacerbate the fine-tuning problem associated with the EWSB scale.   Of course, there are many examples in which natural explanations of the EWSB
scale employ BSM physics with extended Higgs sectors. 
The minimal supersymmetric extension of the SM (MSSM ), which
employs two Higgs doublets, is the most well known example of this type,
but there are numerous other BSM examples as well.
If you give up on naturalness (e.g., vacuum selection), it has been
argued that it may be difficult to accommodate more than
one Higgs doublet at the electroweak scale~\cite{ArkaniHamed:2012kq}.
However, it is possible to construct ``partially natural'' extended Higgs
sectors in which one scalar squared mass parameter is fine-tuned (as in the SM), but
additional scalar mass parameters are related to the EWSB scale by a symmetry~\cite{Draper:2016cag}.

In the rest of this talk, I will take an agnostic approach and entertain the possibility of an extended Higgs sector
without providing a specific theoretical motivation.   I shall focus on the theoretical constraints on extended Higgs
sectors in light of current experimental data (including the fact that the observed Higgs boson is SM-like).   These constraints will
provide an important framework for considering the phenomenology of additional Higgs bosons that could be discovered in future experimentation at the LHC (or at future colliders currently under consideration).

\section{Theoretical implications of a SM-like Higgs boson}

Based on the Run-I LHC Higgs data~\cite{Khachatryan:2016vau}, it is already apparent that the observed Higgs boson is
SM-like. Thus any model of BSM physics, including
models of extended Higgs sectors, must incorporate
this observation.  In models of extended Higgs sectors, a SM-like Higgs
boson can be achieved in a particular limit of the model
called the \textit{alignment limit}~\cite{Gunion:2002zf,Craig:2013hca,Asner:2013psa,Carena:2013ooa,Haber:2013mia}.

Consider an extended Higgs sector with $n$ hypercharge-one Higgs doublets $\Phi_i$ and $m$ additional singlet Higgs fields $\phi_i$.
After minimizing the scalar potential, we assume that only the neutral
scalar fields acquire vevs (in order to preserve electromagnetic charge conservation),
$\vev{\Phi_i^0}=v_i/\sqrt{2}$ and $\vev{\phi_j^0}=x_j$,
where $v^2\equiv \sum_i|v_i|^2=4m_W^2/g^2\simeq (246~{\rm GeV})^2$.
We define new linear combinations, $H_i$,  of the hypercharge-one doublet Higgs fields (this is the so-called \textit{Higgs basis}~\cite{Georgi:1978ri,Branco:1999fs,Davidson:2005cw}).
In particular,
\beq
H_1=\begin{pmatrix} H_1^+ \\ H_1^0\end{pmatrix}=\frac{1}{v}\sum_i v_i^*\Phi_i\,,\qquad\quad
\vev{H_1^0}=v/\sqrt{2}\,,
\eeq
and $H_2, H_3,\ldots, H_n$ are the other mutually orthogonal linear combinations of doublet scalar fields such that $\vev{H_i^0}=0$ (for $i\neq 1$).
That is $H_1^0$ is \textit{aligned} in field space with the direction of the scalar field vev.  
In the alignment limit, $h\equiv\sqrt{2}\,\Re H_1^0-v$ is a mass-eigenstate, and the tree-level couplings of $h$ to itself, to gauge bosons and to fermions are precisely those of the SM Higgs boson.
In general, $\sqrt{2}\,\Re H_1^0-v$ is \textit{not} a mass-eigenstate due to mixing with other neutral scalars.
Thus, the observed Higgs boson is SM-like if at least one of the following two conditions are satisfied:

1. The diagonal squared masses of the other scalar fields are all large compared to the mass of the observed Higgs boson (the so-called \textit{decoupling limit}~\cite{Haber:1989xc,Gunion:2002zf}), and/or

2. The elements of the scalar squared mass matrix that govern the mixing of $\sqrt{2}\,\Re H_1^0 -v$ with other neutral scalars are suppressed.

In the SM, $m_h^2=\lambda v^2$ where $v\simeq 246$~GeV, and $\lambda$ is the Higgs self-coupling [cf.~\eq{V}] which should not be much larger than $O(1)$.  Thus, we expect $m_h\sim O(v)$.  
In extended Higgs sectors, there can be a new mass parameter, $M\gg v$, such that all physical Higgs masses with one exception are of $O(M)$.  The Higgs boson, with $m_h\sim O(v)$, is SM-like due to approximate alignment.  This is the decoupling limit.   After integrating out all the heavy degrees of freedom at the mass scale $M$, one is left with a low-energy effective theory which consists of the SM particles, including a single neutral scalar boson.  This low-energy effective theory is precisely the SM!

The alignment limit is most naturally achieved in the decoupling regime.   However, in this case the additional Higgs boson states are very heavy and may be difficult to observe at the LHC.
In the case of approximate alignment without decoupling\footnote{ In some models, alignment without decoupling can be achieved by a symmetry~\cite{Dev:2014yca,Pilaftsis:2016erj}.  The inert doublet model~\cite{Barbieri:2006dq} is a noteworthy example in which the exact alignment limit is a consequence of a discrete $\mathbb{Z}_2$ symmetry.  In most cases, approximate alignment without decoupling is an accidental region of the model parameter space.}
(due to suppressed scalar mixing), non-SM-like Higgs boson states need not be very heavy and thus are more easily accessible at the LHC.

\section{Examples of extended Higgs sectors near the alignment limit}

\subsection{Extending the SM Higgs sector with a singlet scalar}

The simplest example of an extended Higgs sector adds a real scalar field~$S$.   The
most general renormalizable gauge-invariant scalar potential (subject to a $\mathbb{Z}_2$ symmetry to eliminate linear and cubic terms in $S$) is~\cite{Silveira:1985rk,Burgess:2000yq,Davoudiasl:2004be}
\beq
V(\Phi,S)=-m^2\Phi^\dagger\Phi-\mu^2 S^2+\half\lambda_1(\Phi^\dagger\Phi)^2+\half \lambda_2 S^4
+\lambda_3(\Phi^\dagger\Phi)S^2\,.
\eeq
After minimizing the scalar potential, $\vev{\Phi^0}=v/\sqrt{2}$ and $\vev{S}=x/\sqrt{2}$.  The squared mass matrix of the neutral Higgs bosons is~\cite{Pruna:2013bma,Robens:2015gla}
\beq
\mathcal{M}^2=\begin{pmatrix} \lambda_1 v^2 &  \quad  \lambda_3 vx\\ \lambda_3 vx &  \quad \lambda_2 x^2\end{pmatrix}\,.
\eeq
The corresponding mass eigenstates are $h$ and $H$ with $m_h\leq m_H$.
As discussed in Section 2, an approximate alignment limit can be realized in two different ways:
either  $|\lambda_3|x\ll v$ and/or $x\gg v$.     In the case where
$|\lambda_3|x\ll v$, $h$ is SM-like if $\lambda_1 v^2<\lambda_2 x^2$ and $H$ is SM-like if $\lambda_1 v^2>\lambda_2 x^2$.
In contrast,  $x\gg v$ corresponds to
the \textit{decoupling limit}, where $h$ is SM-like and $m_H\gg m_h$.

The Higgs mass eigenstates are explicitly defined via
\beq
\begin{pmatrix} h \\ H\end{pmatrix}=\begin{pmatrix} \cos\alpha &  \,\,\, -\sin\alpha \\
\sin\alpha & \,\,\, \phm\cos\alpha\end{pmatrix}\begin{pmatrix}\sqrt{2}\,\Re\Phi^0-v \\
\sqrt{2}\,S-x\end{pmatrix}\,,
\eeq
where
\beqa
\lambda_1 v^2&=&m_h^2\cos^2\alpha+m_H^2\sin^2\alpha\,,\\
\lambda_2 x^2&=&m_h^2\sin^2\alpha+m_H^2\cos^2\alpha\,,\\
\lambda_3 xv &=& (m_H^2-m_h^2)\sin\alpha\cos\alpha\,.
\eeqa
The SM-like Higgs boson is approximately given by $\sqrt{2}\,\Re\Phi^0-v$.

If $h$ is SM-like, then $m_h^2\simeq \lambda_1 v^2$ and
\beq
|\sin\alpha|=\frac{|\lambda_3| vx}{\sqrt{(m_H^2-m_h^2)(m_H^2-\lambda_1 v^2)}}\simeq
\frac{|\lambda_3| vx}{m_H^2-m_h^2}\ll1\,,
\eeq
If $H$ is SM-like, then $m_H^2\simeq \lambda_1 v^2$ and
\beq
|\cos\alpha|=\frac{|\lambda_3| vx}{\sqrt{(m_H^2-m_h^2)(\lambda_1 v^2-m_h^2)}}\simeq
\frac{|\lambda_3| vx}{m_H^2-m_h^2}\ll1\,.
\eeq
A phenomenological analysis presented in Ref.~\cite{Robens:2015gla} shows that the allowed parameter regime (consistent with the LHC Higgs data) roughly satisfies $|\sin\alpha|\lsim 0.3$ if $m_H\gsim m_h=125$~GeV and $|\sin\alpha|\gsim 0.9$ if $m_h<m_H=125$~GeV.

\subsection{The two-Higgs doublet model (2HDM)}

Consider the 2HDM with hypercharge-one, doublet fields $\Phi_1$ and~$\Phi_2$~\cite{Gunion:1989we,Branco:2011iw}.
After minimizing the scalar potential, $\vev{\Phi_i^0}=v_i/\sqrt{2}$ (for $i=1,2$),
where $v_1^2+v_2^2\simeq (246~{\rm GeV})^2$ and $\tan\beta\equiv v_2/v_1$.
The Higgs basis fields are defined as,
\beq
H_1=\begin{pmatrix}H_1^+\\ H_1^0\end{pmatrix}\equiv \frac{v_1^* \Phi_1+v_2^*\Phi_2}{v}\,,
\qquad\quad H_2=\begin{pmatrix} H_2^+\\ H_2^0\end{pmatrix}\equiv\frac{-v_2 \Phi_1+v_1\Phi_2}{v}
 \,,
 \eeq
 such that 
  $\vev{H_1^0}=v/\sqrt{2}$ and $\vev{H_2^0}=0$.  The Higgs basis is uniquely defined up to an overall rephasing of the Higgs basis field $H_2$.

In the Higgs basis, the scalar potential is
given by~\cite{Branco:1999fs,Davidson:2005cw,Haber:2006ue}:
\beqa 
V&=& Y_1 H_1^\dagger H_1+ Y_2 H_2^\dagger H_2 +[Y_3
H_1^\dagger H_2+{\rm h.c.}]
+\half Z_1(H_1^\dagger H_1)^2
+\half Z_2(H_2^\dagger H_2)^2
+Z_3(H_1^\dagger H_1)(H_2^\dagger H_2)\nn\\
&&\quad 
+Z_4( H_1^\dagger H_2)(H_2^\dagger H_1)+\left\{\half Z_5 (H_1^\dagger H_2)^2 +\big[Z_6 (H_1^\dagger
H_1) +Z_7 (H_2^\dagger H_2)\big] H_1^\dagger H_2+{\rm
h.c.}\right\}\,,\label{higgspot}
\eeqa
where $Y_1$, $Y_2$ and $Z_1,\ldots,Z_4$ are real, 
whereas $Y_3$, $Z_5$, $Z_6$ and $Z_7$ are potentially complex.
After minimizing the scalar potential, $Y_1=-\half Z_1 v^2$ and $Y_3=-\half Z_6 v^2$.
\clearpage

For simplicity, we consider here 
the case of a CP-conserving scalar potential.\footnote{The more general case in which no scalar basis exists such that all the parameters of
\eq{higgspot} are simultaneously real is treated in Refs.~\cite{Haber:2006ue,Haber:2010bw}.} 
In this case, one can rephase the Higgs basis field $H_2$  such that $\Im Z_5=\Im Z_6=\Im Z_7=0$.
We identify the CP-odd Higgs boson as $A=\sqrt{2}\,\Im~\!H_2^0$, with
$m_A^2=Y_2+\half (Z_3+Z_4-Z_5)v^2$.  After eliminating $Y_2$ in favor of $m_A^2$, the
CP-even Higgs squared mass matrix with respect to the Higgs basis states,
$\{\sqrt{2}\,\Re H^0_1-v$\,,\,$\sqrt{2}\,\Re H^0_2\}$, 
is given by
\beq
\mathcal{M}_H^2=\begin{pmatrix} Z_1 v^2 & \quad Z_6 v^2 \\  Z_6 v^2 & \quad m_A^2+Z_5 v^2\end{pmatrix}\,.
\eeq
The CP-even Higgs bosons are $h$ and $H$ with $m_h\leq m_H$.
The couplings of $\sqrt{2}\,\Re H^0_1-v$ coincide with those of the SM Higgs boson.\footnote{Although the tree-level couplings of $\sqrt{2}\,\Re H_1^0-v$ coincide with those of the SM Higgs boson, the one-loop couplings can differ due to the exchange of non-minimal Higgs states (if not too heavy).  For example, the charged Higgs boson loop interferes with the $W$ and fermion loop contributions to the amplitude for the decay of the SM-like Higgs boson to $\gamma\gamma$ or $\gamma Z$.} Thus,
the alignment limit corresponds to two limiting cases:

1. $|Z_6|\ll 1$.  In this case, $h$ is SM-like if $m_A^2+(Z_5-Z_1)v^2>0$; otherwise, $H$ is SM-like.

2. $m^2_A\gg (Z_1-Z_5)v^2$.  This is the \textit{decoupling limit}; $h$ is SM-like and
$m_A\sim m_H\sim m_{H^\pm}\gg m_h$.

\noindent
In particular, the CP-even mass eigenstates are:
\beq
\begin{pmatrix} H\\ h\end{pmatrix}=\begin{pmatrix} \cbma & \,\,\, -\sbma \\
\sbma & \,\,\,\phantom{-}\cbma\end{pmatrix}\,\begin{pmatrix} \sqrt{2}\,\,\Re H_1^0-v \\ 
\sqrt{2}\,\Re H_2^0
\end{pmatrix}\,,
\eeq
where $\cbma\equiv\cba$ and $\sbma\equiv\sba$ are defined in terms of the angle $\alpha$ that diagonalizes the CP-even Higgs squared mass matrix when expressed in the original basis of scalar fields, $\{\sqrt{2}\,\Re\Phi_1^0-v_1\,,\,\sqrt{2}\,\Re\Phi_2^0-v_2\}$, and $\tan\beta\equiv v_2/v_1$.
Since the SM-like Higgs boson is approximately $\sqrt{2}\,\Re H_1^0-v$, it follows that
$h$ is SM-like if $|\cbma|\ll 1$, and
$H$ is SM-like if $|\sbma|\ll 1$.

The approximate alignment limit can be  derived more explicitly as follows.
The CP-even Higgs squared mass matrix yields~\cite{Haber:2015pua,Bernon:2015qea}
\beqa
Z_1 v^2&=&\mhl^2 s^2_{\beta-\alpha}+\mhh^2 c^2_{\beta-\alpha}\,,\nn\\
Z_6 v^2&=&(\mhl^2-\mhh^2)\sbma\cbma\,,\nn \\
Z_5 v^2&=&\mhh^2 s^2_{\beta-\alpha}+\mhl^2 c^2_{\beta-\alpha}-m_A^2\,.\nn
\eeqa
If $h$ is SM-like, then $m_h^2\simeq Z_1 v^2$ and
\beq \label{eqcbma}
|\cbma|=\frac{|Z_6| v^2}{\sqrt{(\mhh^2-\mhl^2)(\mhh^2-Z_1 v^2)}}\simeq \frac{|Z_6| v^2}{m_H^2-m_h^2}\ll 1\,.
\eeq
The decoupling limit is realized when $m_H\gg m_h$.   In contrast, alignment without decoupling requires that $|Z_6|\ll 1$ and $m_H\sim O(v)$.
If $H$ is SM-like, then $m_H^2\simeq Z_1 v^2$ and~\cite{Bernon:2015wef}
\beq
|\sbma|=\frac{|Z_6| v^2}{\sqrt{(\mhh^2-\mhl^2)(Z_1 v^2-\mhl^2)}}\simeq \frac{|Z_6| v^2}{m_H^2-m_h^2}\ll 1\,,
\eeq
which can only be achieved if $|Z_6|\ll 1$.  In particular a SM-like $H$ can only arise in the limit of alignment without decoupling.
\clearpage

So far, we have not yet discussed the couplings of the Higgs bosons to fermions.
In the $\Phi_1$--$\Phi_2$ basis, the 2HDM Higgs-quark Yukawa Lagrangian is~\cite{Haber:2010bw}:
\beq
-\mathscr{L}_{\rm Y}=\overline U_L \Phi_{i}^{0\,*}{{h^U_i}} \ur -\anti
D_L K^\dagger\Phi_{i}^- {{h^U_i}}\ur
+\overline U_L K\Phi_i^+{{h^{D\,\dagger}_{i}}} \dr
+\overline D_L\Phi_i^0 {{h^{D\,\dagger}_{i}}}\dr+{\rm h.c.}\,,
\eeq
where $K$ is the CKM mixing matrix, $h^{U,D}$ are $3\times 3$ Yukawa coupling matrices,
and there is an implicit sum over~$i=1,2$.  
Unlike in the SM, the diagonalization of the quark masses does not automatically diagonalize the neutral-Higgs--quark Yukawa coupling matrices.  Hence, the general 2HDM possesses
tree-level Higgs-mediated flavor-changing neutral currents (FCNCs) which 
are generically too large and thus inconsistent with experimental data.
In order to \textit{naturally} eliminate tree-level Higgs-mediated FCNC~\cite{Glashow:1976nt,Paschos:1976ay},
one can impose a discrete symmetry to restrict the 
structure of $\mathscr{L}_{\rm Y}$ .Two different
choices for how the discrete symmetry acts on the quarks then yield:
Type-\Rmnum{1} Yukawa couplings~\cite{Haber:1978jt,Hall:1981bc} if $h_1^U=h_1^D=0$,
and Type-\Rmnum{2} Yukawa couplings~\cite{Donoghue:1978cj,Hall:1981bc} if $h_1^U=h_2^D=0$.
(Similar considerations can also be applied to the Higgs-lepton Yukawa couplings.)
\begin{table}[b!]
\centering
\begin{tabular}{|c||c|c|}\hline
Higgs interaction & 2HDM coupling & approach to alignment limit \\
\hline
$hVV$ & $\sbma$ & $1-\half c^2_{\beta-\alpha}$ \\[6pt]
$hhh$ & *
& $1+2(Z_6/Z_1)\cbma$  \\[6pt]
$hH^+H^-$ & * &
$\tfrac{1}{3}\left[(Z_3/Z_1)+(Z_7/Z_1)\cbma\right]$\\[6pt]
$Hhh$ & * &
$-Z_6/Z_1+\bigl[1-\tfrac23(Z_{345}/Z_1)\bigr]c_{\beta-\alpha}$\\[6pt]
$hhhh$ & *
& $1+3(Z_6/Z_1)\cbma$  \\[6pt]
$h\overline{D}D$ & $\sbma\mathds{1}+\cbma\rho^D_R$
& $\mathds{1}+\cbma\rho^D_R$\\[6pt]
$h\overline{U}U$ & $\sbma\mathds{1}+\cbma\rho^U_R$
& $\mathds{1}+\cbma\rho^U_R$\\[6pt] \hline
\end{tabular}
\caption{The 2HDM couplings of the SM-like Higgs boson $h$ normalized to
those of the SM Higgs boson, in the approach to the
alignment limit.  The $hH^+H^-$ and $Hhh$ couplings are normalized to
the SM $hhh$ coupling~\cite{Gunion:2002zf,Bernon:2015qea} (where $Z_{345}\equiv Z_3+Z_4+Z_5$).
The scalar Higgs potential is taken to be CP-conserving.
For the Higgs couplings to fermions,
$D$ is a column vector of three down-type fermion fields
(either down-type quarks or charged leptons)
and $U$ is a column vector of three up-type quark fields.   For  Type-I Yukawa couplings,  $\rho_R^D=\rho_R^U=\mathds{1}\cot\beta$,
and for Type-II Yukawa couplings, $\rho_R^D=-\mathds{1}\tan\beta$ and $\rho_R^U=\mathds{1}\cot\beta$~\cite{Asner:2013psa}.
In the third column above, the first non-trivial correction to alignment is exhibited.  Finally, complete expressions for the entries marked with a * can be found in Refs.~\cite{Haber:2006ue,Haber:2010bw}.}
\end{table}

It is straightforward to work out the Higgs couplings in the approximate alignment limit.   Some examples are provided in Table 1 in the case where $h$ is SM-like with a CP-conserving scalar potential and Type-I or II Yukawa couplings.  In the third column of Table 1, the first non-trivial corrections to the alignment limit are presented.  Note that these corrections are correlated, and the approach to decoupling is governed by $\cbma$.  Thus, any deviations from SM-like behavior of the observed Higgs boson can provide important clues to the structure of the extended Higgs sector.  The phenomenology of the 2HDM in the approximate alignment limit and its implications for future LHC experimental studies have recently been elucidated in Refs.~\cite{Bernon:2015qea,Bernon:2015wef} in the case where $h$ or $H$, respectively, is identified as the observed Higgs boson of mass 125 GeV.

\subsection{The MSSM Higgs sector}

The MSSM Higgs sector is a CP-conserving Type-II 2HDM.  The dimension-four terms of the scalar potential are constrained by supersymmetry.  In particular, the scalar potential parameters (at tree level) in the Higgs basis are determined by the electroweak gauge couplings~\cite{Haber:2006ue},
\beqa
Z_1&=&Z_2=\quarter(g^2+g^{\prime\,2}) c_{2\beta}^2\,,\qquad Z_5=\quarter(g^2+g^{\prime\,2})s_{2\beta}^2\,,\qquad
Z_7=-Z_6=\quarter(g^2+g^{\prime\,2}) s_{2\beta}c_{2\beta}\,,\nn \\
Z_3&=&Z_5+\quarter(g^2-g^{\prime\,2})\,,\qquad\quad\,\,\,\,
Z_4=Z_5-\half g^2\,,\label{zsusy} 
\eeqa
in a convention where $\tan\beta\geq 0$,
where $c_{2\beta}\equiv\cos 2\beta$ and $s_{2\beta}\equiv \sin 2\beta$.   It then follows from \eq{eqcbma} that,
\beq
\cos^2(\beta-\alpha)=\frac{m_Z^4 \,s^2_{2\beta} c^2_{2\beta}}{(m_H^2-m_h^2)(m_H^2-m_Z^2 c^2_{2\beta})}\,.
\eeq
The decoupling limit is achieved when $m_H\gg m_h$ as expected.   Exact alignment without decoupling is (naively) possible
at tree-level when $Z_6=0$, which yields $\sin 4\beta= 0$ and $m_h^2=Z_1 v^2=m_Z^2 c^2_{2\beta}$.   However, 
these results are inconsistent with the observed Higgs mass of 125~GeV.

It is well known that radiative corrections can significantly modify the tree-level Higgs properties~\cite{Draper:2016pys}.
In particular, consider the limit  
where $m_h$, $m_A$, $m_H$, $m_{H^\pm}\ll M_S$, where $M_S^2\equiv m_{\tilde t_1}m_{\tilde t_2}$ is the product of top squark masses.   In this case, one can formally integrate out the squarks and generate a low-energy effective 2HDM Lagrangian (which is no longer of the tree-level MSSM form).
At one-loop, the dominant contributions to the effective $Z_1$ and $Z_6$ parameters are given by~\cite{Carena:2014nza}
\footnote{CP-violating phases , which could appear in the MSSM parameters such as $\mu$ and $A_t$, are neglected.}
 \beqa
Z_1 v^2 &=& m_Z^2 c^2_{2\beta}+\frac{3v^2 s_\beta^4 h_t^4}{8\pi^2}\left[\ln\left(\frac{M_S^2}{m_t^2}\right)+\frac{X_t^2}{M_S^2}\left(1-\frac{X_t^2}{12M_S^2}\right)\right]\,,\label{zeeone}\\[8pt]
Z_6 v^2&=& -s_{2\beta}\left\{m_Z^2 c_{2\beta}-\frac{3v^2 s_\beta^2  h_t^4}{16\pi^2}\biggl[\ln\left(\frac{M_S^2}{m_t^2}\right)+\frac{X_t(X_t+Y_t)}{2M_S^2}-\frac{X_t^3 Y_t}{12 M_S^4}\biggr]\right\}\,,\label{zeesix}
\eeqa
where $s_\beta\equiv\sin\beta$, $h_t$ is the top quark Yukawa coupling, $X_t\equiv A_t-\mu\cot\beta$ and $Y_t\equiv A_t+\mu\tan\beta$.  

\begin{figure}[t!]
\centering
\includegraphics[width=0.32\textwidth]{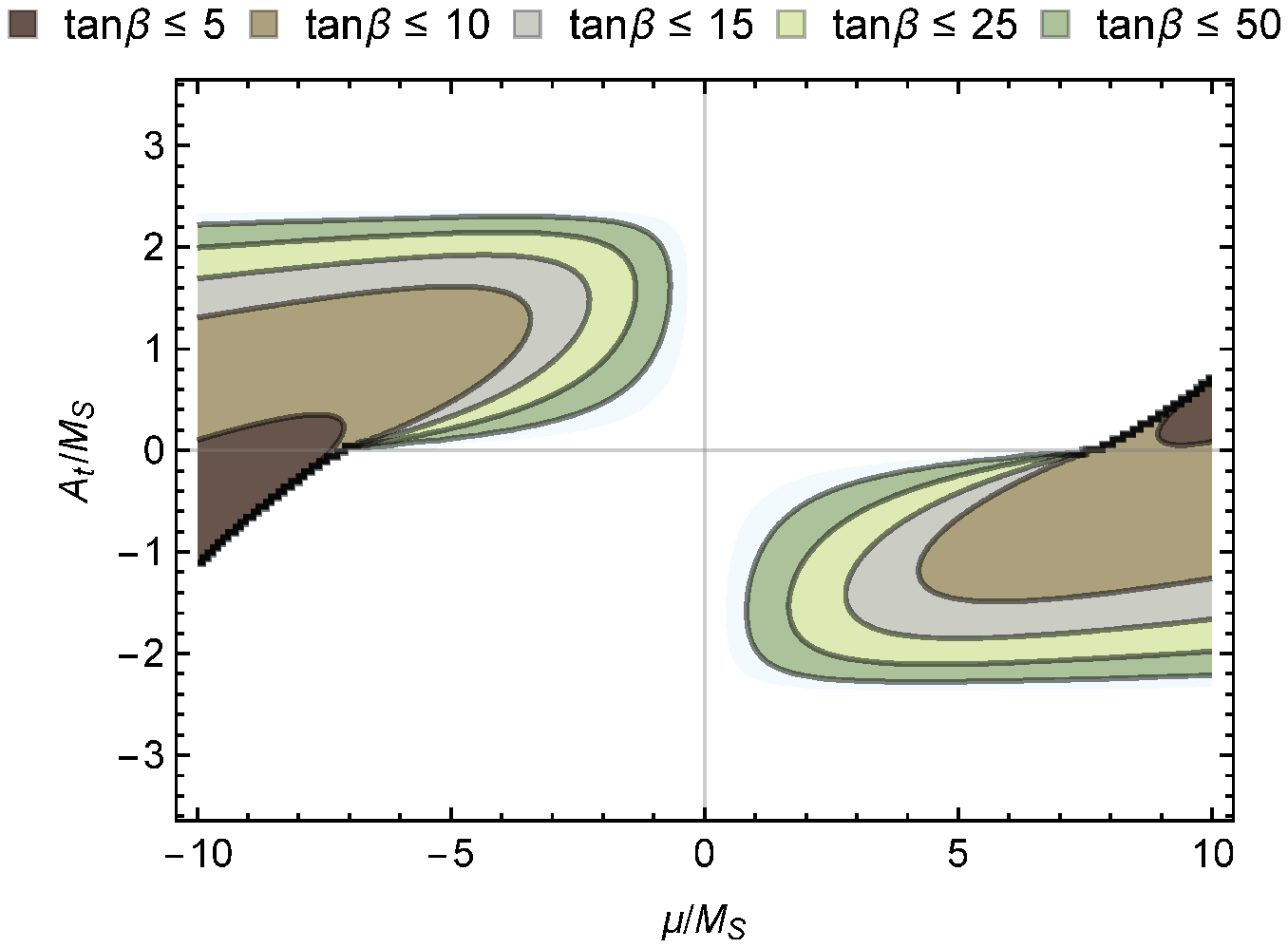}
\includegraphics[width=0.32\textwidth]{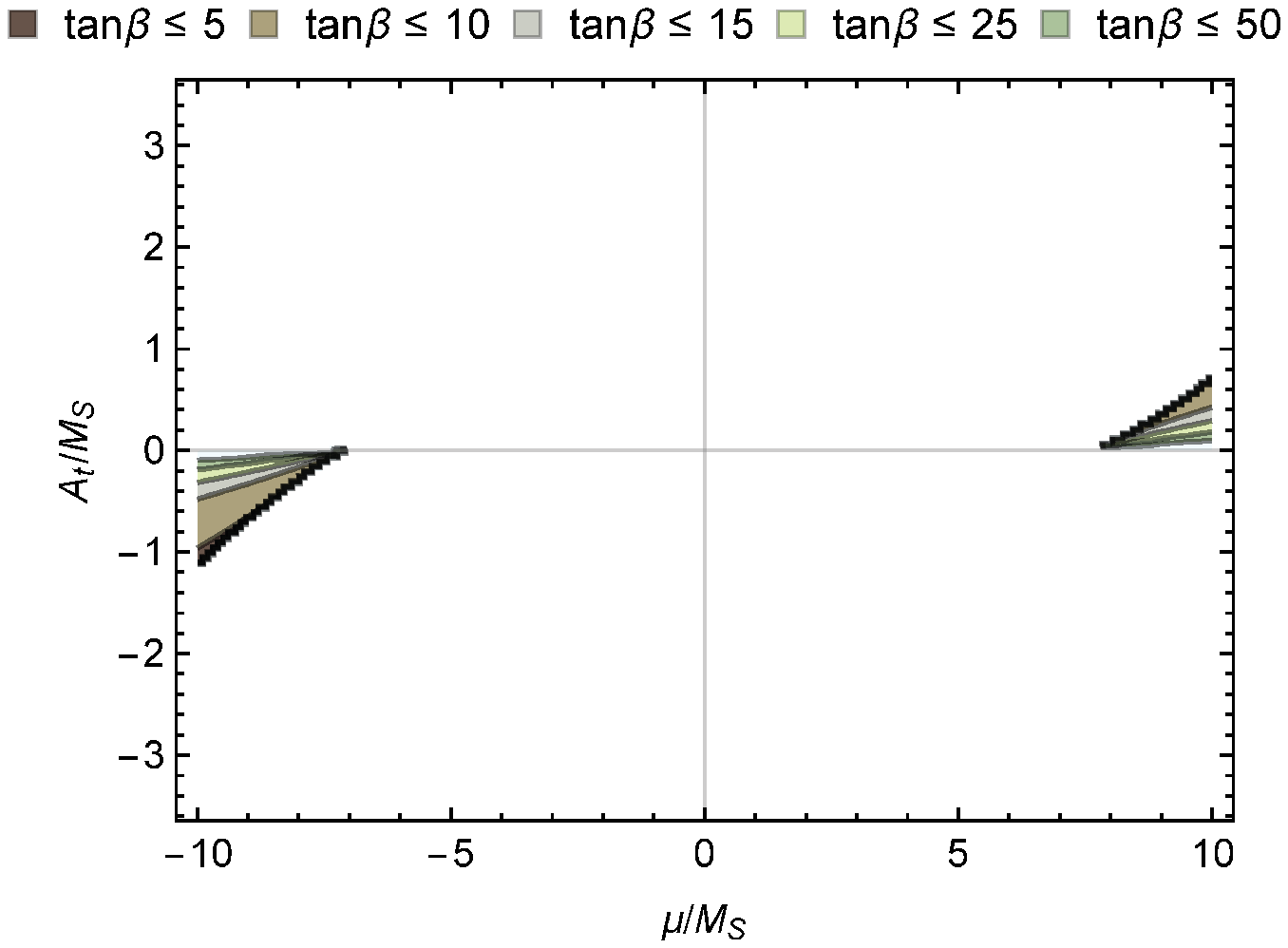}\
\includegraphics[width=0.32\textwidth]{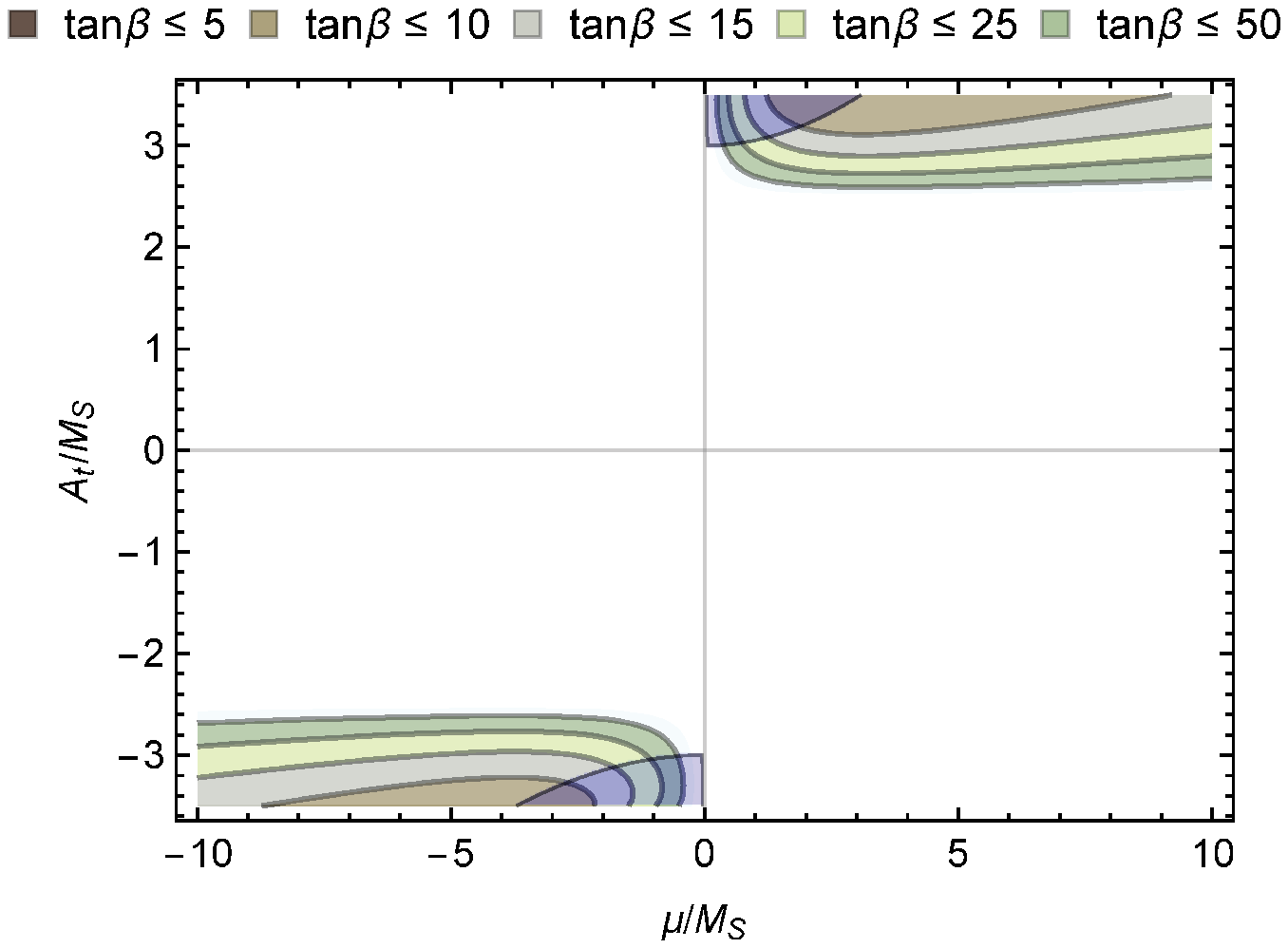}
\caption{Contours of $\tan\beta$ corresponding to exact alignment, $Z_6= 0$,
in the $(\mu/M_S, A_t/M_S)$ plane.
$Z_1$ is adjusted to give the correct Higgs mass.
The leading one-loop and two-loop corrections of $O(\alpha_s h_t^2)$ to $Z_1$ and $Z_6$ have been included.  Taking the three panels together, one can immediately discern the regions of zero, one, two and
three values of $\tan\beta$ in which exact alignment is realized.
Taken from Ref.~\cite{Bechtle:2016kui}.\label{tb}}
\end{figure}

Note that
$m_h^2\simeq Z_1 v^2$ is consistent with $m_h\simeq 125$~GeV for suitable choices for $M_S$ and $X_t$.
Exact alignment (i.e., $Z_6=0$) can now be achieved due to an accidental cancellation between tree-level and loop contributions~\cite{Carena:2014nza},
\beq \label{oneloopalign}
m_Z^2 c_{2\beta}=\frac{3v^2 s_\beta^2  h_t^4}{16\pi^2}\biggl[\ln\left(\frac{M_S^2}{m_t^2}\right)+\frac{X_t(X_t+Y_t)}{2M_S^2}-\frac{X_t^3 Y_t}{12 M_S^4}\biggr]\,.
\eeq
One can manipulate \eq{oneloopalign} into a 7th order polynomial equation in $\tan\beta$.
The alignment condition is then achieved by (numerically) solving
this equation for positive real solutions of $\tan\beta$.  Following a recipe provided by Refs.~\cite{Haber:1996fp,Carena:2000dp}, one can 
further improve \eqs{zeeone}{zeesix} to include the leading
two-loop corrections of $O(\alpha_s h_t^2)$ by replacing $h_t$ with  $h_t(\lambda)$, where $\lambda\equiv \bigl[m_t(m_t)M_S\bigr]^{1/2}$ in the one-loop leading log contributions and $\lambda\equiv M_S$ in the leading threshold corrections.  Imposing $Z_6=0$ now leads to a
11th order polynomial equation in $\tan\beta$ that can be solved numerically.  Three positive solutions are exhibited in Fig.~\ref{tb} as a function of $\mu/M_S$ and $A_t/M_S$~\cite{Bechtle:2016kui}.

In light of the SM-like nature of the observed Higgs boson, the ATLAS Collaboration concluded that $m_A\gsim 370$~GeV~\cite{Aad:2015pla}.
However, this analysis failed to consider the possibility of approximate alignment without decoupling~\cite{Carena:2014nza}, which can be achieved in certain regions of the MSSM parameter space~\cite{Bechtle:2016kui}.  The direct searches for $H$ and $A$ (decaying into $\tau^+\tau^-$) by the ATLAS and CMS Collaborations in the mass region from 200--370~GeV provide no constraints for values of $\tan\beta\lsim 8$--$10$~\cite{ATLAS:2016fpj,CMS:2016rjp}.   In this parameter regime, approximate alignment is still possible for suitable choices of $\mu/M_S$ and $A_t/M_S$.  A recent pMSSM parameter scan~\cite{Bechtle:2016kui}, which takes into account the observed Higgs data, direct searches for $H$ and $A$, indirect constraints from heavy flavor physics, and supersymmetric particle searches, finds that values of $m_A$ as low as 200~GeV are within 2$\sigma$ of the best fit point obtained by a global likelihood analysis, under the assumption that $h$ is  the observed Higgs boson of mass 125 GeV.\footnote{The same analysis also yields an allowed parameter regime where $H$ is the observed Higgs boson of mass 125~GeV.} 

Alignment without decoupling can also be achieved in the NMSSM (where an additional Higgs singlet superfield is added to the MSSM).   For further details, see Ref.~\cite{Carena:2015moc}.

\subsection{Beyond Higgs singlets and doublets}

If one considers a scalar sector with triplet Higgs fields, then one must also include additional Higgs multiplets in such a way that the electroweak $\rho$-parameter is approximately equal to 1.    Georgi and Machacek constructed a model in which $\rho=1$ at tree-level due to a well chosen scalar potential that respects the custodial symmetry~\cite{Georgi:1985nv}.  Their model contains a complex $Y=1$ doublet, a complex $Y=2$ triplet and a real $Y=0$ singlet.  After minimizing the scalar potential, there is a doublet vev,
$v_\phi$, and a common triplet vev, $v_\chi$, with $v^2\equiv v_\phi^2+8v_\chi^2\simeq (246~{\rm GeV})^2$.

The physical scalars make up custodial SU(2)  multiplets:
a 5-plet of states ($H_5^{\pm\pm}$, $H_5^\pm$ and $H_5^0$) with common mass $m_5$, a triplet 
($H_3^\pm$, $H_3^0$) with common mass $m_3$, and custodial singlets that mix with squared mass matrix~\cite{Hartling:2014zca}
\beq
\mathcal{M}^2=\begin{pmatrix} Z_{11}v_\phi^2 & \quad v_\phi v_\chi(Z_{12}-2\sqrt{3}\, m_3^2/v^2) \\
v_\phi v_\chi(Z_{12}-2\sqrt{3}\,m_3^2/v^2) & \quad \tfrac{3}{2} m_3^2-\half m_5^2+v_\chi^2(Z_{22}-12m_3^2/v^2)\end{pmatrix}\,,
\eeq
where the $Z_{ij}$ depend on the dimensionless quartic couplings.
The custodial singlet CP-even Higgs bosons are $h$ and $H$ with $m_h\leq m_H$.
An approximate alignment limit can be realized in two different ways.  First, if $v_\chi\ll v$, then $h$ is SM-like if $Z_{11}v^2<\tfrac{3}{2} m_3^2-\half m_5^2$; otherwise, $H$ is SM-like.  Alternatively, in
the decoupling limit, $h$ is SM-like and
$m_H\simeq m_3\simeq m_5\gg m_h$~\cite{Hartling:2014zca}.

One interesting feature of  the Georgi-Machacek model is that the existence of doubly-charged Higgs bosons modifies the unitarity sum rule~\cite{Gunion:1990kf},
\beq \label{sumrule}
\sum_i g^2_{h_iW^+W^-}=g^2 m_W^2+\sum_k|g_{H^{++}_k W^- W^-}|^2\,,
\eeq
where the sum is taken over all CP-even Higgs bosons of the model.  The presence on the last term on the right hand side of \eq{sumrule} means that individual $h_i VV$ couplings can exceed the corresponding SM Higgs coupling to $VV$.
It is convenient to write
$
c_H\equiv\cos\theta_H={v_\phi}/(v_\phi^2+8v_\chi^2)^{1/2}\,,
$
and $s_H\equiv\sin\theta_H$.   Then, the following couplings are noteworthy~\cite{Gunion:1989ci}:
\beqa
&&H_1^0 W^+ W^-:\quad g c_H m_W\,,\qquad\qquad\,\, \,   H_1^{\prime\,0}W^+ W^-:\quad \sqrt{8/3}gm_W s_H\,,\nn\\
&&H_5^0 W^+ W^-:\quad \sqrt{1/3}gm_W s_H\,,\qquad H_5^{++}W^-W^-:\quad \sqrt{2}gm_W s_H\,,\nn
\eeqa
where $H_1^0$ and $H_1^{\prime\,0}$ are the custodial singlet interaction eigenstates.  Among the four Higgs states that couple to $WW$, whose couplings are listed above, $H_1^{\prime\,0}$, $H_5^0$ and $H_5^{++}$ have no coupling to fermions, whereas the $H_1^0 f\bar{f}$ coupling is
$-gm_q/(2m_W c_H)$.

In general $H_1^0$ and $H_1^{\prime\,0}$ can mix.  In the absence of $H_1^0$--$H_1^{\prime\,0}$ mixing, $c_H=1$ corresponds to the alignment limit. But consider the strange case of $s_H=\sqrt{3/8}$, where the
$H_1^{\prime\,0}$ coupling to $W^+ W^-$ matches that of the SM.  Nevertheless, this does not saturate the $HWW$ sum rule!  Moreover, it is possible that the $H_1^{\prime\,0}W^+ W^-$ coupling is \textit{larger} than the SM value of $gm_W$, without violating the sum rule given by \eq{sumrule}.  Including $H_1^0$--$H_1^{\prime\,0}$ mixing allows for even more baroque scenarios that are not possible in a multi-doublet extension of the SM Higgs sector.

\section{Conclusions}

Given the non-minimal nature of the observed spectrum of fundamental fermions and gauge bosons, it would be remarkable if the Higgs boson were a solo act.  Thus, the search for additional scalars that exist in an extended Higgs sector will be an important enterprise in the experimental program at the LHC and at any future collider facility.

The current Higgs data strongly suggest that the observed Higgs boson is SM-like.  
This already places a strong constraint on the theoretical structure of any non-minimal Higgs sector.  
In particular, the alignment limit, in which the mass eigenstate corresponding to the observed Higgs boson is aligned with the direction (in field space) of the scalar doublet vev, must be a good approximation.   The simplest way to achieve the alignment limit is in the case where all additional Higgs scalars are significantly heavier than the observed Higgs boson (corresponding to the decoupling limit).   But, we have also argued for the possibility of the alignment limit without decoupling if the mixing between the SM Higgs boson and the additional neutral Higgs scalars is suppressed, in which case all Higgs scalars may be light [of $O(v)$] and thus more accessible to LHC searches.

Finally, as the Higgs data become more precise, deviations from SM properties of the Higgs boson may eventually be observed.   Indeed, 
departures from the alignment limit encode critical information that can provide important clues for the structure of the non-minimal Higgs sector.  Pursuing Higgs physics into the future by theorists and experimentalists is likely to lead to profound insights into
the fundamental theory of particles and their interactions.

\section*{Acknowledgments}

This work is supported in part by the U.S. Department of Energy grant number DE-SC0010107.   Travel support and the hospitality of Rikard Enberg, Arnaud Ferrari  and Uppsala University are also gratefully acknowledged.

\end{document}